\documentclass[twocolumn,amsmath,amssymb,prl,superscriptaddress,a4paper,floatfix,showkeys]{revtex4}
\usepackage{graphicx}
\usepackage{mathptmx}      % use Times fonts if available on your TeX system

\newcommand\MT{Maxwell's theory}
\newcommand\QT{quantum theory}

\begin{document}

\title{Event-by-event simulation of nonclassical effects in two-photon interference experiments\footnote{
Invited paper presented at FQMT11.\\
Accepted for publication in Physica Scripta 27 June 2012.}}

\author{Kristel Michielsen}
\email{k.michielsen@fz-juelich.de}           %  \\
\affiliation{%
Institute for Advanced Simulation, J\"ulich Supercomputing Centre,
Forschungzentrum J\"ulich, D-52425 J\"ulich, Germany
}%
\affiliation{%
RWTH Aachen University, D-52056 Aachen, Germany
}%
\author{Fengping Jin}
\email{f.jin@fz-juelich.de}           %  \\
\affiliation{%
Institute for Advanced Simulation, J\"ulich Supercomputing Centre,
Forschungzentrum J\"ulich, D-52425 J\"ulich, Germany
}%
\author{M. Delina}
\email{m.m.delina@rug.nl}
\affiliation{%
Department of Applied Physics,
Zernike Institute for Advanced Materials,
University of Groningen, Nijenborgh 4, NL-9747 AG Groningen, The Netherlands
}%
\affiliation{%
Physics Department,
Faculty of Mathematics and Natural Science,
State University of Jakarta,
Jl.Permuda No 10, Jakarta 13220, Indonesia
}%
\author{Hans De Raedt}
\email{h.a.de.raedt@rug.nl}
\affiliation{%
Department of Applied Physics,
Zernike Institute for Advanced Materials,
University of Groningen, Nijenborgh 4, NL-9747 AG Groningen, The Netherlands
}%

\begin{abstract}
A corpuscular simulation model for second-order intensity interference phenomena is discussed.
It is shown that both the visibility ${\cal V}=1/2$ predicted for two-photon interference experiments with two independent sources
and the visibility ${\cal V}=1$ predicted for two-photon interference experiments with a parametric down-conversion source
can be explained in terms of a locally causal, modular, adaptive, corpuscular, classical (non-Hamiltonian) dynamical system.
Hence, there is no need to invoke quantum theory to explain the so-called nonclassical effects in the interference of signal and
idler photons in parametric-down conversion. A revision of the commonly accepted criterion
of the nonclassical nature of light is needed.
\end{abstract}

\date{\today}

\maketitle

\section{Introduction}

In classical optics, interference is known to be a phenomenon in which two waves are superimposed, resulting in a wave
with bigger or smaller amplitude. Observed for the first time in Young's two-slit experiment in 1803~\cite{YOUN04}, it played an
important role in the general acceptance of the wave character of light.
In quantum theory, interference in the two-slit experiment with electrons, large molecules, photons and other so-called quantum particles
is considered to demonstrate the wave-particle duality of these quantum particles.
In fact, according to Feynman the observation that the interference pattern in the two-slit experiment with electrons is built
up detection event by detection event is a phenomenon which is ``impossible, absolutely impossible, to explain in any classical way,
and which has in it the heart of quantum mechanics''~\cite{FEYN65}. He referred to the interference of
single electrons as ``the only mystery'' of quantum mechanics~\cite{FEYN65}.

In general, a classical optical interference experiment consists of several classical light sources (not necessarily primary sources)
and several detectors which measure the resulting light intensity at various positions. Adding equipment
that accumulates the time average of the product of the detector signals allows for the measurement of the second and higher order intensity
correlations. In quantum optics, the sources are replaced by single photon sources (the primary source commonly said to create
single photons or $N$-photon entangled states with $N\ge 2$) and single photon detectors. A coincidence circuit is added
to the experimental setup to measure coincidences in the photon counts.

In this paper we limit the number of sources and the number of detectors to two.
Interference is then characterized by the dependence of the resulting light intensity or of the second order intensity correlations
on certain phase shifts.
The Hanbury Brown-Twiss (HBT) effect was one of the first observations that demonstrated interference in the intensity-intensity correlation
functions~\cite{HANB56}. HBT showed that under conditions for which the usual two-beam interference fringes measured by each of the two detectors vanish,
the correlated intensities of the two-detectors can still show interference fringes.
For two completely independent sources, be it classical light sources or single photon sources,
the visibility of this second-order intensity interference has an upperbound of 1/2~\cite{MAND99}.
For primary sources producing correlated photon pairs, such as parametric down-converting sources,
the two sources in an HBT-type of experiment can no longer be considered to be independent.
In that case the two sources are considered to emit exactly one photon of the correlated pair simultaneously.
Such sources provide a $100\%$ visibility of the second-order intensity correlation, exceeding the $50\%$ limit which is
a commonly accepted criterion of nonclassicality~\cite{MAND99}.
The first experiment devoted to demonstrate nonclassical second-order intensity interference effects in the absence of first order intensity interference
is probably the Ghosh-Mandel two-photon interference experiment of 1987~\cite{GHOS87}.
%Although one year earlier the effect was also obeserved in experiments conducted to measure the absolute quantum efficiencies
%of photodetectors with simultaneously produced pairs of photons with parametric down-conversion~\cite{BOWM86}.
However, the effect is not limited to photons.
Second-order intensity interference effects have also been observed in two-atom interference experiments~\cite{YASU96,OTTL05,SCHE05,JELT07}
in which an expanding cloud of cooled atoms acts as a source, multi-channel
plate(s) detect the arrival and position of a particle, and time-coincidence techniques
are employed to obtain the two-particle correlations.
Also in Hanbury Brown-Twiss type of experiments with electrons second-order intensity interference effects have been
observed~\cite{OLIV99,KIES02}.
As well intensity interference in the two-slit experiment as second-order intensity interference in Hanbury Brown and Twiss-type of
experiments is attributed to the dual wave-particle character of the quantum particles.

In previous work~\cite{RAED05d,RAED05b,MICH11a,ZHAO08b,MICH10a,JIN10c,JIN10b,JIN10a,MICH12a}
we have demonstrated, using an event-based corpuscular model,
that interference is not necessarily a signature of the presence of waves
of some kind but can also appear as the collective result of particles which at any time do not
directly interact with each other.
In general, the event-based approach deals with the fact that experiments yield definite results, such as
for example the individual detector clicks that build up an interference pattern.
We call these definite results ``events''.
Instead of trying to fit the existence of these events
in some formal, mathematical theory, in the event-based approach the paradigm is changed
by directly searching for the rules that transform events
into other events and, which by repeated application,
yield frequency distributions of events that
agree with those predicted by classical wave or quantum theory.
Obviously, such rules cannot be derived from quantum theory or,
as a matter of fact, of any theory that is probabilistic in nature
simply because these theories do not entail a procedure (= algorithm)
to produce events themselves.

The event-based approach has successfully been
used to perform discrete-event simulations of the single beam splitter and
Mach-Zehnder interferometer experiment
of Grangier {\sl et al.}~\cite{GRAN86} (see Refs.~\cite{RAED05d,RAED05b,MICH11a}),
Wheeler's delayed choice experiment of Jacques {\sl et al.}~\cite{JACQ07}
(see Refs.~\cite{ZHAO08b,MICH10a,MICH11a}),
the quantum eraser experiment of Schwindt {\sl et al.}~\cite{SCHW99} (see Ref.~\cite{JIN10c,MICH11a}),
double-slit and two-beam single-photon interference experiments and the single-photon interference experiment with
a Fresnel biprism of Jacques {\sl et al.}~\cite{JACQ05} (see Ref.~\cite{JIN10b,MICH11a}),
quantum cryptography protocols (see Ref.~\cite{ZHAO08a}),
the Hanbury Brown-Twiss experiment of Agafonov {\sl et al.}~\cite{AGAF08} (see Ref.~\cite{JIN10a,MICH11a}),
universal quantum computation (see Ref.~\cite{RAED05c,MICH05}),
Einstein-Podolsky-Rosen-Bohm-type of experiments of Aspect {\sl et al.}~\cite{ASPE82a,ASPE82b}
and Weihs {\sl et al.}~\cite{WEIH98} (see Refs.~\cite{RAED06c,RAED07a,RAED07b,RAED07c,RAED07d,ZHAO08,MICH11a}),
and the propagation of electromagnetic plane waves through homogeneous thin films and stratified media (see Ref.~\cite{TRIE11,MICH11a}).
An extensive review of the simulation method and its applications is given in Ref.~\cite{MICH11a}.
Proposals for single-particle experiments to test specific aspects of the event-based approach are discussed in Refs.~\cite{JIN10b,MICH12a}

In this paper, we demonstrate that the second-order intensity interference with visibility 1/2 in a HBT experiment with two
independent single photon sources and with visibility 1 in the Ghosh-Mandel experiment can be entirely explained in terms of
an event-based model, that is in terms of a locally causal, modular, adaptive, classical (non-Hamiltonian) dynamical system.
Hence, there is no need to invoke quantum theory to explain the observations and the commonly accepted criterion
of the nonclassical nature of light needs to be revised.

\begin{figure}[t]
\begin{center}
\includegraphics[width=8cm ]{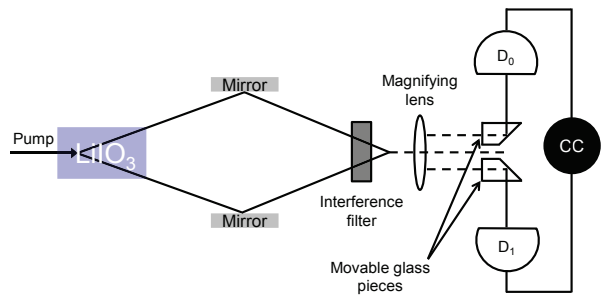}
\caption{%
Diagram of the Ghosh--Mandel interference experiment~\cite{GHOS87}.
A source emits pairs of single-photons through spontaneous down-conversion
in a LiIO$_3$ crystal.
These photons leave the source in different directions.
Mirrors redirect the photons to the interference filter and a lens.
The two beams overlap at a distance of about $1\,$m from the crystal.
The resulting image is magnified by a lens and two movable glass pieces
are used to collect and redirect the photons to
the single-photon detectors $D_0$ and $D_1$,
the signals of which are fed into a coincidence counter CC.
}%
\label{fig.1}
\end{center}
\end{figure}

\section{Second-order intensity interference}

\begin{figure}[t]
\begin{center}
\includegraphics[width=8cm]{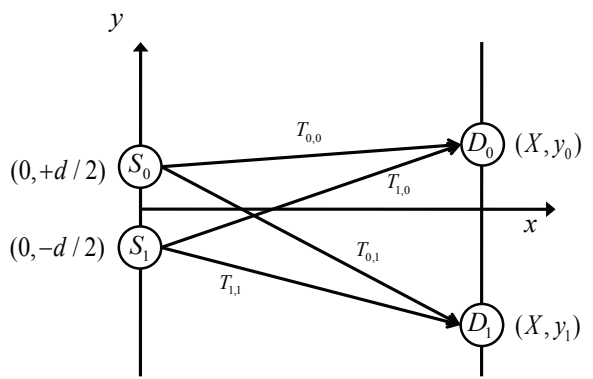}
\caption{Schematic diagram of the Ghosh--Mandel experiment.
Single photons emitted from point sources $S_0$ and $S_1$ positioned at the $y$ axis and separated by a center-to-center
distance $d$ are registered by two detectors $D_0$ and $D_1$ positioned on a line at a distance $X$ from the $y$ axis.
The time of flight for each of the four possible paths from source $S_m$ to detector $D_n$
is denoted by $T_{m,n}$ where $m,n=0,1$.
}%
\label{exphbt}
\end{center}
\end{figure}

In the context of the Ghosh--Mandel experiment, see Fig.~\ref{fig.1}, we may view the two mirrors as the two sources
that produce two overlapping beams of photons.
Hence, conceptually, this experiment can be simplified as shown in Fig.~\ref{exphbt}, which is
the schematic diagram of a HBT experiment~\cite{MAND99}.

A HBT experiment is nothing but a two-beam experiment with two independent sources and two detectors.
The two sources are positioned along the $y$-axis and are separated by a center-to-center distance $d$.
The two detectors are placed on a line at a distance $X$ from the $y$-axis.
Assume that source $S_m$ ($m=0,1$) emits coherent light of frequency $f$
and produces a wave with amplitude $A_m e^{i\phi_m}$ ($A_m$ and $\phi_m$ real).
For simplicity of presentation, we assume that $A_0=A_1=A$.
According to \MT, the total wave amplitude $B_n$ on detector $n$ is
\begin{equation}
B_n = A \left(e^{i(\phi_0+2\pi fT_{0,n})} + e^{i(\phi_1+2\pi fT_{1,n})}\right)
,
\end{equation}
where the time of flight for each of the four possible paths from source $S_m$ to detector $D_n$
is denoted by $T_{m,n}$ where $m,n=0,1$.
The light intensity $I_n = |B_n|^2$ on detector $D_n$ is given by
\begin{equation}
I_n = 2A^2\left\{ 1+ \cos \left[\phi_0-\phi_1+2\pi f(T_{0,n}-T_{1,n})\right]\right\}
.
\label{hbt0}
\end{equation}
If the phase difference $\phi_0-\phi_1$ in Eq.~(\ref{hbt0}) is fixed, the usual two-beam  (first-order) interference fringes are observed.

The essence of the HBT experiment is that if the phase difference $\phi_0-\phi_1$ is a random variable
(uniformly distributed over the interval $[0,2\pi[$) as a function of observation time,
these first-order interference fringes vanish because
\begin{equation}
\langle I_n\rangle = 2A^2,
\label{hbt1}
\end{equation}
where $\langle . \rangle$ denotes the average over the variables $\phi_0$ and $\phi_1$.
However, the average of the product of the intensities $I_0$ and $I_1$ is given by
\begin{equation}
\langle I_0I_1\rangle = 4A^4\left( 1+ \frac{1}{2}\cos 2\pi f \Delta T\right)
,
\label{hbt2}
\end{equation}
where $\Delta T =(T_{0,0}-T_{1,0})-(T_{0,1}-T_{1,1})$.
Accordingly, the intensity-intensity correlation Eq.~(\ref{hbt2}) exhibits second-order interference fringes,
a manifestation of the so-called HBT effect.
From Eqs.~(\ref{hbt1}) and (\ref{hbt2}), it follows that the visibility of the signal $I$, defined by
\begin{equation}
{\cal V}= \frac{\max(I)-\min(I)}{\max(I)+\min(I)}
,
\label{hbt2a}
\end{equation}
is given by ${\cal V}=0$ and ${\cal V}=1/2$ for the first-order and second-order intensity interference, respectively.

Treating the electromagnetic field as a collection of bosons changes Eq.~(\ref{hbt2}) into~\cite{MAND99}
\begin{equation}
\langle I_0I_1\rangle^{\mathrm{bosons}} = 4A^4\left( 1+ \cos 2\pi f \Delta T\right)
.
\label{hbt2b}
\end{equation}
%For fermionic fields we have
%\begin{equation}
%\langle I_0I_1\rangle^{\mathrm{fermions}} = A^4\left( 1 - \cos 2\pi f \Delta T\right)
%.
%\label{hbt2c}
%\end{equation}
Clearly, for bosons, the visibility of the second-order intensity interference is ${\cal V}=1$.

Considering the situation in which the two independent sources $S_0$ and $S_1$
are replaced by sources that emit simultaneously exactly one photon of a correlated photon pair
emitted by a parametric down-conversion source
gives a similar expression for the average of the product of the intensities $I_0$ and $I_1$ as given
by Eq.(~\ref{hbt2b}). Hence, also in this case ${\cal V}=1$ for the second-order intensity interference.

In the two-beam experiment interference appears in its most pure form because
the phenomenon of diffraction is absent.
If we assume that the detectors cannot communicate with each other,
that there is no direct communication between the particles involved
and that it is indeed true that individual pairs of particles build up the interference pattern one by one,
just looking at Fig.~\ref{exphbt} leads to the logically unescapable
conclusion that the interference can only be due to the internal operation of the detector~\cite{PFlE67}.
Detectors that simply count the incoming photons are not sufficient to explain the appearance of an interference pattern
and apart from the detectors there is nothing else that can cause the interference pattern to appear.
We now discuss an event-based model of a detector that can cope with this problem~\cite{MICH11a}.

\section{Simulation model}\label{sec2}

The model discussed in this paper builds on our earlier work~\cite{RAED05d,RAED05b,RAED05c,MICH11a}.
In short, in our simulation approach, a photon is viewed as a messenger that carries a message
and material is regarded as a message processor.
Evidently, the messenger itself can be thought of as a particle.
For the present purpose, it suffices to encode in the message, the time of flight of the particle.
The interaction of the photons with material translates into a processing unit receiving,
manipulating and sending out messages.
Note that we explicitly prohibit two particles from communicating directly
and that interference results from the processing of individual particles only~\cite{RAED05d,RAED05b,RAED05c,JIN10a,JIN10b,MICH11a}.

We now explicitly describe the model, that is
we specify the message carried by the messengers, the algorithm for simulating
a detector ( = processing unit), and the simulation procedure itself.

\textbf{Messenger:} The messenger can be regarded as a particle which travels with velocity $c$ in the direction $\mathbf{q}/q$.
Each messenger carries with it a harmonic oscillator which vibrates with frequency $f$.
It may be tempting to view the messenger with its message as a plane wave
with wave vector $\mathbf{q}$, the oscillator being one of the two electric field
components in the plane orthogonal to $\mathbf{q}$.
However, this analogy is superfluous and should not be stretched too far.
As there is no communication/interaction between the messengers
there is no wave equation (i.e.~no partial differential equation)
that enforces a relation between the messages carried by different messengers.
Indeed, the oscillator carried by a messenger never interacts with the oscillator
of another messenger, hence the motion of these pairs of oscillators is not governed by a wave equation.
Naively, one might imagine the oscillators tracing out a wavy pattern as they travel through space.
However, as there is no relation between the times at which the messengers leave the source,
it is impossible to characterize all these traces by a field that
depends on one set of space-time coordinates, as required for a wave theory.
It is convenient (though not essential) to represent the
message, that is the oscillator,
by a two-dimensional unit vector ${\mathbf y}=\left( \cos \psi,\sin \psi\right) $ where $\psi=2\pi f t+\delta$.
Here, $t$ is the time of flight of the particle and $\delta$ is a phase shift.
Pictorially, the message is nothing but a representation of the hand of a clock which
rotates with period $1/f$ and is running ahead by a time related to the phase $\delta$.
A processing unit has access to this data and may use the messenger's internal clock
to determine how long it took for the messenger to reach the unit.

\begin{figure}[t]
\begin{center}
\includegraphics[width=8cm]{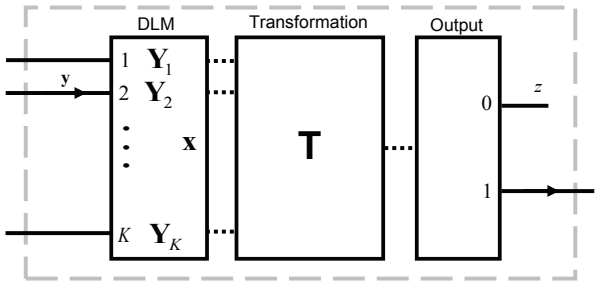}
\caption{%
Diagram of the event-based detector model defined by Eqs.~(\ref{det0}) -- (\ref{det3}).
The detection unit consists of an input stage, which is a deterministic learning machine (DLM), a transformation stage and an output stage.
The input stage has $K$ input channels at which a message ${\mathbf y}$, being a two-component vector, can arrive,
$K$ corresponding internal registers ${\mathbf Y}_k$ in which the incoming message can be stored and one
internal $K$-component vector ${\mathbf x}$, responsible for the learning.
The transformation stage generates a message ${\mathbf T}$, a two-component vector, based on all information available in the input stage.
The output stage takes the message ${\mathbf T}$ as input and generates an output signal $z$ representing a ``click'' or ``no click''
on output channel 0 or 1, respectively.
The detection unit processes one message at a time.
The solid lines indicate the input and output channels of the processing unit and the
dashed lines indicate the data flow within the processing unit.
}%
\label{figdetector}
\end{center}
\end{figure}

\textbf{Source:} A source creates a messenger % (particle)
with its phase $\delta$ set to some randomly chosen value.
Initially its time of flight $t$ is zero as it  is determined by the arrival of the messenger at a processing unit.
A pseudo-random number determines to which detector the messenger travels.

\textbf{Single-photon detector:} In reality, photon detection is the result of a complicated
interplay of different physical processes~\cite{HADF09}.

In essence, a light detector consists of material that absorbs light.
The electric charges that result from the absorption process are then amplified,
chemically in the case of a photographic plate or electronically in the case of photodiodes or photomultipliers.
In the case of photomultipliers or photodiodes,
once a photon has been absorbed (and its energy ``dissipated'' in the detector material)
an amplification mechanism (which requires external power/energy) generates
an electric current (provided by an external current source)~\cite{GARR09,HADF09}.
The resulting signal is compared with a threshold that is set
by the experimenter and the photon is said to have been detected
if the signal exceeds this threshold~\cite{GARR09,HADF09}.
In the case of photographic plates, the chemical process that occurs when photons
are absorbed and the subsequent chemical reactions that
renders visible the image serve similar purposes.

Photon detectors, such as a photographic plate of CCD arrays, consist of many identical detection units
each having a predefined spatial window in which they can detect photons.
In what follows, each of these identical detection units will be referred to as a detector.
By construction, these detector units operate completely independently from and also do not communicate with each other.

An event-based model for the detector cannot be ``derived'' from  \QT\
simply because \QT\ has nothing to say about individual events
but predicts the frequencies of their observation only~\cite{HOME97}.
Therefore, any model for the detector that operates on the level of single events
must necessarily appear as ``ad hoc'' from the viewpoint of  \QT.
The event-based detector model that we employ in this paper should not be regarded
as a realistic model for say, a photomultiplier or a photographic plate and the
chemical process that renders the image.
In the spirit of Occam's razor, the very simple event-based model captures the salient features
of ideal (i.e. 100\% efficient)
single-photon detectors.

The key element of the event-by-event
approach is a processing unit that is adaptive, that is it can learn
from the messengers that arrive at its input ports~\cite{RAED05d,RAED05b,MICH11a}.
The diagram of an event-based detection unit is depicted in Fig.~\ref{figdetector}.
It consists of an input stage called deterministic learning machine (DLM)~\cite{RAED05d,RAED05b},
a transformation stage, and an output stage.
The processing unit should act as a detector for individual messengers which may come
from several different directions. Therefore, as can be seen from
the schematic diagram depicted in Fig.~\ref{figdetector} this processing unit
has $K$ input ports, a parameter that allows the machine to resolve $K$ different directions.

\textit{Input stage:} Representing the arrival of a messenger at port $1\le k\le K$ by
the vector ${\bf v}=(v_1,\ldots,v_K)^T$ with $v_i=\delta_{i,k}$ ($i=1,\ldots K$)
the internal vector is updated according to the rule
\begin{equation}
\mathbf{x} \leftarrow \gamma \mathbf{x} + (1-\gamma) \mathbf{v}
,
\label{det0}
\end{equation}
where $\mathbf{x}=(x_1,\ldots,x_{K})^T$, $\sum_{k=1}^K x_k=1$, and $0\le\gamma<1$.
The elements of the incoming message $\mathbf{y}$ are written in internal register $\mathbf{Y}_k$
\begin{equation}
\mathbf{Y}_k\leftarrow\mathbf{y}
,
\label{det1}
\end{equation}
while all the other $\mathbf{Y}_{i}$ ($i\not=k$) registers remain unchanged.
Thus, each time a messenger arrives at one of the input ports, say $k$,
the DLM updates all the elements of the internal vector $\mathbf{x}$,
overwrites the data in the register $\mathbf{Y}_{k}$
while the content of all other $\mathbf{Y}$ registers remains the same.

\textit{Transformation stage}: The output message generated by the transformation stage is
\begin{equation}
\mathbf{T}=\mathbf{x}\cdot\mathbf{Y}=\sum_{k=1}^K x_k \mathbf{Y}_k
,
\label{det2}
\end{equation}
which is a two-component vector. Note that $|\mathbf{T}|\le1$.

\textit{Output stage}: As in all previous event-based models for the optical components,
the output stage generates a binary output signal $z=0,1$
but the output message does not represent a photon:
It represents a ``no click'' or ``click'' if $z=0$ or $z=1$, respectively.
To implement this functionality, we define
\begin{equation}
z = \Theta(|\mathbf{T}|^2- {\cal R})
,
\label{det3}
\end{equation}
where $\Theta(.)$ is the unit step function and $0\leq {\cal R} <1$
are uniform pseudo-random numbers (which are different for each event).
The parameter $0\le \gamma<1$ can be used to control the operational mode of the unit.
From Eq.~(\ref{det3}) it follows that the frequency of $z=1$ events depends
on the length of the internal vector $\mathbf{T}$.

Note that in contrast to experiment, in a simulation, we could register both the $z=0$ and $z=1$ events.
Then the sum of the $z=0$ and $z=1$ events is equal to the number of input messages.
In real experiments, only $z=1$ events are taken as evidence that a photon has been detected.
Therefore, we define the total detector count by
\begin{equation}
N_{\mathrm{count}}=\sum^{N}_{l=1}z_l,
\label{det4}
\end{equation}
where $N$ is the number of messages received and $l$ labels the events.
In other words, $N_{\mathrm{count}}$ is the total number of one's generated by the detector unit.

Comparing the number of ad hoc assumptions and unknown functions that enter quantum theoretical treatments
of photon detectors~\cite{GARR09} with the two parameters $\gamma$ and $K$ of the event-based detector model,
the latter has the virtue of being extremely simple while providing a description of
the detection process at the level of detail, the single events, which in any case is outside the scope of \QT.

\textbf{Simulation procedure:} Before the simulation starts we set ${\bf x}=(1,0,\ldots ,0)^T$
and we use pseudo-random numbers ${\cal R}$ to set $\mathbf{Y}_{k}=(\cos2\pi{\cal R},\sin2\pi{\cal R})$ for $k=1,\ldots,K$.
Next, we generate $N_{\mathrm{tot}}$ pairs of messengers, send them to the detectors, determine the detector count $N_{\mathrm{count}}$ at $D_0$ and $D_1$
and count the coincidences. In the simulation always two messengers travel to the detectors, one generated at source $S_0$ and one at
source $S_1$. Hence, once a pair of messengers is generated a detector can generate no click, one click or two clicks.
Only when both detectors generate a click the coincidence count $N_{\mathrm {coincindence}}$ is enhanced by one.

\section{Simulation results}

\subsection{Detection efficiency}

The efficiency of the detector model is determined by simulating an experiment
that measures the detector efficiency, which for a single-photon detector is defined
as the overall probability of registering a count if a photon arrives at the detector~\cite{HADF09}.
In such an experiment a point source emitting single particles is placed far away from a single detector.
As all particles that reach the detector have the same time of flight (to a very good approximation), all the
particles that arrive at the detector will carry nearly the same message $\mathbf{y}$ which is encoding the time of flight.
Furthermore, they arrive at the same input port, say $q$.
As a result ${\mathbf x}$ (see Eq.~(\ref{det0})) rapidly converges to the vector with
$x_i\rightarrow \delta_{i,q}$ and, as $\mathbf{y}$ is a unit vector, we have $|\mathbf{T}|\approx 1$,
implying that the detector clicks almost every time a photon arrives.
Thus, for our detector model, the detection efficiency as defined for real detectors~\cite{HADF09} is very close
to 100\% (results not shown).

\subsection{Hanbury Brown-Twiss experiment}\label{HBT}

\begin{figure}[t]
\begin{center}
\includegraphics[width=8cm]{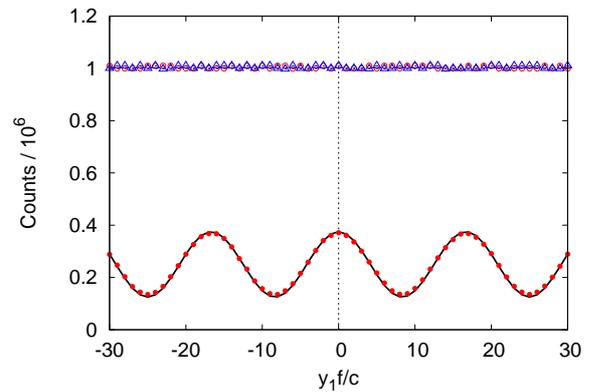}
\caption{%
Simulation data of the single-particle and two-particle counts
for the HBT experiment depicted in Fig.~\ref{exphbt}.
Red open circles (Blue open triangles): results for the counts $N_{\mathrm {count}}$ of detector $D_0$ ($D_1$),
showing that there is no second-order intensity interference.
Red closed circles: results for the coincidence counts $N_{\mathrm{coincidence}}$.
The dashed and solid lines represent the theoretical predictions $N_{tot}/2$
and Eq.~(\ref{hbt4}) for the single detector and coincidence counts, respectively.
Simulation parameters: $N_{\mathrm{tot}}=2\times 10^6$ events per $y_1f/c$-value,
$N_{\mathrm{F}}=50$, $X=100000c/f$, $d=2000c/f$, $\gamma=0.99$ and $K=2$.
}
\label{simhbt0}
\end{center}
\end{figure}

In Fig.~\ref{simhbt0} we present the simulation results for the HBT experiment depicted in Fig.~\ref{exphbt}.
For simplicity, we have put detector $D_0$ at $(X,0)$ and plot the single detector and
coincidence counts as a function of the $y$-position of detector $D_1$.
In each simulation step, both sources $S_0$ and $S_1$ create a messenger % (particle)
with some randomly chosen phase being the only initial content of the messages $\mathbf{y}_{m}$ ($m=0,1$).
The phases are kept fixed for $N_{\mathrm{F}}$ successive pairs of messengers.
The total number of emitted pairs is denoted by $N_{\mathrm {tot}}$.
Two pseudo-random numbers are used to determine whether the messengers travel to detector $D_0$ or $D_1$.
The time of flight for the messenger travelling from source $S_m$ to detector $D_n$ is given by
\begin{equation}
T_{m,n}=\frac{\sqrt{X^2+((1-2m)d/2 - y_n)^2}}{c}
,
\label{hbt4a}
\end{equation}
where $m,n=0,1$. The time of flight $T_{m,n}$ is added to the message $\mathbf{y}_{m}$ before the message is processed by the
corresponding detector $D_n$. The messages are the only input to the event-based model.
As Fig.~\ref{simhbt0} shows, averaging over the randomness in the initial messages (random phases) wipes
out all interference fringes in the single-detector counts, in agreement with \MT.
We find that the number of single-detector counts $N_{\mathrm {count}}$ fluctuates around $N_{\mathrm{tot}}/2$, as expected from wave theory.
Similarly, the data for the coincidence counts are in excellent agreement with the theoretical prediction for
the simulation model
\begin{equation}
N_{\mathrm{coincidence}} = \frac{N_{\mathrm{tot}}}{8} \left(1+ \frac{1}{2}\cos 2\pi f \Delta T\right)
,
\label{hbt4}
\end{equation}
and, disregarding the prefactor $N_{\mathrm{tot}}/8$, also in qualitative agreement with the predictions of wave theory.

For simplicity, we have confined the above presentation to the case of a definite polarization.
Simulations with randomly varying polarization (results not shown) are also in concert
with \MT.

\subsection{Ghosh-Mandel experiment}

From Eq.~(\ref{hbt4}), it follows that the visibility of the interference fringes, defined by
\begin{equation}
{\cal V}= \frac{\max(N_{\mathrm{coincidence}})-\min(N_{\mathrm{coincidence}})}{\max(N_{\mathrm{coincidence}})+\min(N_{\mathrm{coincidence}})}
,
\label{hbt5}
\end{equation}
cannot exceed 1/2.
It seems commonly accepted that the visibility of a two-photon interference experiment exceeding 1/2
is a signature of the nonclassical nature of light.

As two-photon interference experiments, such as the Gosh-Mandel experiment~\cite{GHOS87},
employ time-coincidence to measure the intensity-intensity correlations,
it is quite natural to expect that a model that purports to explain the observations accounts for the time delay
that occurs between the time at which a particle arrives at a detector and the actual click of that detector.
In \QT, time is not an observable and can therefore not be computed within the theory proper.
Hence there is no way that these time delays, which are being measured, can be accounted for by  \QT.
Consequently, any phenomenon that depends on these time delays must find an explanation
outside the realm of  \QT\ (as it is formulated to date).

It is straightforward to add a time-delay mechanism to the event-based model of the detector.
For simplicity, let us assume that the time delay for the detector click is given by
\begin{equation}
t_{\mathrm{delay}}=T_{m,n}-T_{\mathrm{max}}(1-|\mathbf{T}|^2)^h\ln{\cal R}
,
\label{hbt6}
\end{equation}
where $0<{\cal R}<1$ is a pseudo-random number, and $\mathbf{T}$ is given by Eq.~(\ref{det2}).
The time scale $T_{\mathrm{max}}$ and the exponent $h$ are
free parameters of the time-delay model.
Note that $t_{\mathrm{delay}}-T_{m,n}$ is a pseudo-random variable
drawn from an exponential distribution with mean $T_{\mathrm{max}}(1-|\mathbf{T}|^2)^h$.
Coincidences are counted by comparing the difference between the delay times
of detectors $D_0$ and $D_1$ with a time window $W$.

\begin{figure}[t]
\begin{center}
\includegraphics[width=8cm]{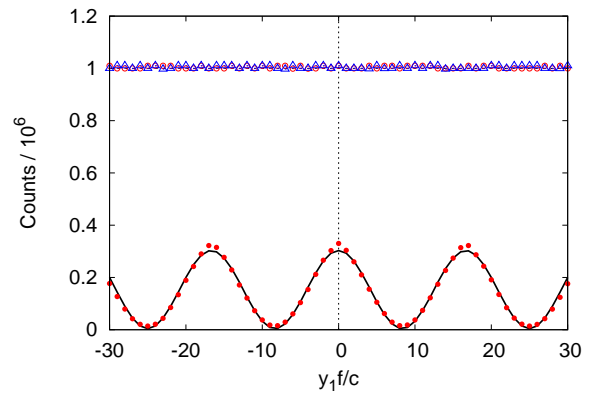}
\caption{%
Simulation data of the single-particle and two-particle counts
for the HBT experiment depicted in Fig.~\ref{exphbt}, generated
by the same event-based that produced the data of Fig.~\ref{simhbt0}
extended with the time-delay model Eq.~(\ref{hbt6}).
Simulation parameters: $T_{\mathrm{max}}/f=1000$, $W/f=1$, $h=8$.
The dashed and solid lines are least-square fits
to $a'_2N_{tot}$ and $a'_4N_{tot}(1+b'_4\cos2\pi f\Delta t)$
for the single detector and coincidence counts, $N_{\mathrm{count}}$ and $N_{\mathrm{coincidence}}$, respectively.
The values of the fitting parameters are
$a'_2=0.502$, $a'_4=0.077$ and $b'_4=0.974$.
%Red open circles: EBCM results for the counts of detector $D_0$.
%Blue open triangles: EBCM results for the counts of detector $D_1$.
%Red closed circles: EBCM results for the coincidence counts.
%The dashed and solid lines are least-square fits of the predictions of wave theory
%to the EBCM data for the single detector and coincidence counts, respectively.
%Simulation parameters: $N_{\mathrm{tot}}=2\times 10^6$ events per $y_1f/c$-value,
%$N_{\mathrm{F}}=50$, $X=100000c/f$, $d=2000c/f$, $\gamma=0.99$, $K_p=2$,
%$W=2/f$, $T_{\mathrm{max}}=1000/f$ and $h=8$.
}
\label{simhbt1}
\end{center}
\end{figure}

From the simulation results presented in Fig.~\ref{simhbt1},
it is clear that by taking into account that there are fluctuations
in the time delay that depend on the time of flight and the internal state
of the detector, the visibility changes from ${\cal V}=1/2$ to ${\cal V}\approx1$.
The simulation data is represented (very) well by $N'_{\mathrm{count}}\approx N_{\mathrm{tot}}/2$ and
\begin{equation}
N'_{\mathrm{coincidence}} \approx a'_4 N_{tot}\left(1+ \cos 2\pi f \Delta T\right)
,
\label{hbt10}
\end{equation}
where the prime indicates that the model incoporates the time-delay mechanism
and $a^\prime_4$ is a fitting parameter which depends on the details
of the time-delay mechanism.
As expected, the use of a narrow time window leads to a signicant reduction (by a factor $a'_4=0.077$) of the total coincidence count.
These results demonstrates that a purely classical corpuscular model of a two-photon interference experiment
can yield visibilities that are close to one.
Hence, the commonly accepted criterion of the nonclassical nature of light needs to be revised.

The time delay model Eq.~(\ref{hbt6}) is perhaps one of the simplest that yield interesting results but
it is by no means unique and can only be scrutinized on the basis of accurate experimental data
which, unfortunately, do not seem to be available thus far.

\subsection{Bosons}

\begin{figure}[t]
\begin{center}
\includegraphics[width=8cm]{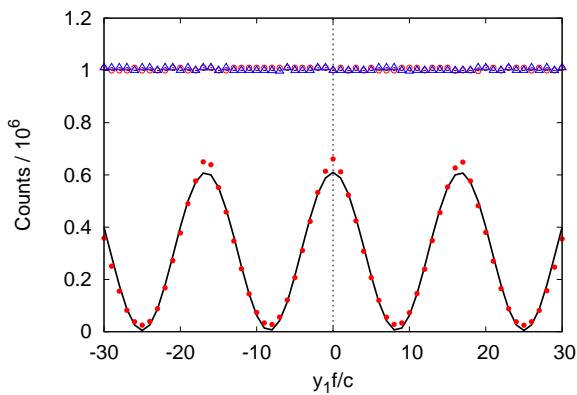}
\caption{%
Same as Fig.~\ref{simhbt1} except that the two sources never send their
particles to the same detector, mimicking bosons (see text).
The values of the fitting parameters are $a''_2=0.502$, $a''_4=0.154$ and $b''_4=0.985$.
%Red open circles: EBCM results for the counts of detector $D_0$.
%Blue open triangles: EBCM results for the counts of detector $D_1$.
%Red closed circles: EBCM results for the coincidence counts.
%The dashed and solid lines are least-square fits of the predictions of wave theory
%to the EBCM data for the single detector and coincidence counts, respectively.
%Simulation parameters: $N_{\mathrm{tot}}=2\times 10^6$ events per $y_1f/c$-value,
%$N_{\mathrm{F}}=50$, $X=100000c/f$, $d=2000c/f$, $\gamma=0.99$, $N_p=2$,
%$W=2/f$, $T_{\mathrm{max}}=1000/f$ and $h=8$.
}
\label{simhbt2}
\end{center}
\end{figure}

If we exclude the possibility that the two sources send their particles
to the same detector, the event-based approach produces results
that are reminiscent of the quantum theoretical description in terms of bosons.
In Fig.~\ref{simhbt2}, we present the results of such a simulation,
using the same model parameters as those used to produce the results
of Fig.~\ref{simhbt1}.
From Figs.~\ref{simhbt1} and \ref{simhbt2}, it is clear
that the maximum amplitude of the two-particle interference signal
of the latter is two times larger than that of the former (the ``classical'' case),
as expected for bosons.
The simulation data is represented (very) well by $N''_{\mathrm{count}}\approx N_{\mathrm{tot}}/2$
\begin{equation}
N''_{\mathrm{coincidence}} \approx a''_4 N_{tot} \left(1+ \cos 2\pi f \Delta T\right)
,
\label{hbt11}
\end{equation}
where the double prime indicates that the model incoporates the time-delay mechanism
and that the possibility that the two sources send their particles
to the same detector has been excluded.

\subsection{Non-monochromatic sources}

All the results presented above have been obtained by assuming that the
beams of particles are strictly monochromatic, meaning that the frequency
$f$ of the oscillators carried by the particles is fixed.
A more realistic simulation of
the pairs of photons created by the parametric down-conversion process
requires that the frequencies $f_1$ and $f_2$ of the messages carried by the pair of particles
satisfy energy conservation, meaning that $f_1+f_2=f_0$ where $f_0$ is the frequency
of the pump beam~\cite{RUBI92,SHIH93,SHIH94,GARR09}.
It is straightforward to draw the
frequencies $f_1$ (and therefore $f_2=f_0-f_1$) from a specified
distribution, such as a Lorentzian~\cite{RUBI92,GARR09}.
In the simulation, each created particle pair would then correspond
to one message characterized by a frequency $f_1$
and another one by frequency $f_2$.
The detectors simply sum all the contributions (taking into account
the differences in the factors $f_m T_{m,n}$), resulting in a
reduction of the visibility, just as in the wave mechanical picture.

%\begin{figure}[t]
%\begin{center}
%\includegraphics[width=8cm]{figure6.eps}
%\caption{Schematic diagram of a HBT experiment.
%Single photons emitted from point sources $S_0$ and $S_1$ are registered by two detectors $D_0$ and $D_1$.
%The time of flight for each of the four possible paths from source $S_m$ to detector $D_n$
%is denoted by $T_{m,n}$ where $m,n=0,1$.
%}%
%\label{exphbtbosons}
%\end{center}
%\end{figure}
%

%\begin{figure}[t]
%\begin{center}
%\includegraphics[width=8cm]{figure8.eps}
%\caption{%
%Same as Fig.~\ref{simhbt1} except that the source never sends two
%particles to the same detector and that the messages carried by the particles are
%modified to mimick fermions (see text).
%The values of the fitting parameters are $a_2=0.502$, $a_4=0.154$ and $b_4=-0.987$.
%%Red open circles: EBCM results for the counts of detector $D_0$.
%%Blue open triangles: EBCM results for the counts of detector $D_1$.
%%Red closed circles: EBCM results for the coincidence counts.
%%The dashed and solid lines are least-square fits of the predictions of wave theory
%%to the EBCM data for the single detector and coincidence counts, respectively.
%%Simulation parameters: $N_{\mathrm{tot}}=2\times 10^6$ events per $y_1f/c$-value,
%%$N_{\mathrm{F}}=50$, $X=100000c/f$, $d=2000c/f$, $\gamma=0.99$, $N_p=2$,
%%$W=2/f$, $T_{\mathrm{max}}=1000/f$ and $h=8$.
%}
%\label{simhbt3}
%\end{center}
%\end{figure}

\section{Conclusion}

We have shown that the so-called nonclassical effects observed in two-photon interference experiments
with a parametric down-conversion source can be explained in terms of a locally causal, modular, adaptive, corpuscular, classical
(non-Hamiltonian) dynamical system.
The high visibility, ${\cal V}=1$, in this type of experiment is commonly considered as a signature of two-photon light, in contrast
to the visibility ${\cal V}=1/2$ obtained in a similar experiment with a classical light source.
On the other hand, according to Ref.~\cite{AGAF08}, the existence of high-visibility interference
in the third and higher orders in the intensity cannot be considered as a signature of three- or four-photon interference, because
high-visibility interference is also observed in Hanbury Brown-Twiss type interference experiments with classical light.
Hence, although the case of second-order intensity interference seemed to be different from the higher orders, we have demonstrated that
also for the second order intensity interference the value of the visibility cannot be used to say anything about the quantum
character of the source. As well the interference experiment with a classical light source as the interference experiment with the
parametric down-conversion source can be explained entirely in terms of a classical corpuscular model.

Elsewhere, we have shown that third order intensity interference in a Hanbury Brown-Twiss type of experiment with two
sources emitting uncorrelated single photons can be modeled by an event-based model as well~\cite{JIN10a}.
Simulation of an interference experiment with a three-photon source is left for future research.

\bibliography{c:/d/papers/epr11,c:/d/papers/neutrons,c:/d/papers/hbt}   % name your BibTeX data base

\begin{thebibliography}{45}
\expandafter\ifx\csname natexlab\endcsname\relax\def\natexlab#1{#1}\fi
\expandafter\ifx\csname bibnamefont\endcsname\relax
  \def\bibnamefont#1{#1}\fi
\expandafter\ifx\csname bibfnamefont\endcsname\relax
  \def\bibfnamefont#1{#1}\fi
\expandafter\ifx\csname citenamefont\endcsname\relax
  \def\citenamefont#1{#1}\fi
\expandafter\ifx\csname url\endcsname\relax
  \def\url#1{\texttt{#1}}\fi
\expandafter\ifx\csname urlprefix\endcsname\relax\def\urlprefix{URL }\fi
\providecommand{\bibinfo}[2]{#2}
\providecommand{\eprint}[2][]{\url{#2}}

\bibitem[{\citenamefont{Young}(1804)}]{YOUN04}
\bibinfo{author}{\bibfnamefont{T.}~\bibnamefont{Young}},
  \bibinfo{journal}{Phil. Trans. R. Soc. Lond.} \textbf{\bibinfo{volume}{94}},
  \bibinfo{pages}{1} (\bibinfo{year}{1804}).

\bibitem[{\citenamefont{Feynman et~al.}(1965)\citenamefont{Feynman, Leighton,
  and Sands}}]{FEYN65}
\bibinfo{author}{\bibfnamefont{R.~P.} \bibnamefont{Feynman}},
  \bibinfo{author}{\bibfnamefont{R.~B.} \bibnamefont{Leighton}},
  \bibnamefont{and} \bibinfo{author}{\bibfnamefont{M.}~\bibnamefont{Sands}},
  \emph{\bibinfo{title}{The Feynman Lectures on Physics, Vol. 3}}
  (\bibinfo{publisher}{Addison-Wesley}, \bibinfo{address}{Reading MA},
  \bibinfo{year}{1965}).

\bibitem[{\citenamefont{Brown and Twiss}(1956)}]{HANB56}
\bibinfo{author}{\bibfnamefont{R.~H.} \bibnamefont{Brown}} \bibnamefont{and}
  \bibinfo{author}{\bibfnamefont{R.}~\bibnamefont{Twiss}},
  \bibinfo{journal}{Nature} \textbf{\bibinfo{volume}{177}}, \bibinfo{pages}{27
  } (\bibinfo{year}{1956}).

\bibitem[{\citenamefont{Mandel}(1999)}]{MAND99}
\bibinfo{author}{\bibfnamefont{L.}~\bibnamefont{Mandel}},
  \bibinfo{journal}{Rev. Mod. Phys.} \textbf{\bibinfo{volume}{71}},
  \bibinfo{pages}{S274 } (\bibinfo{year}{1999}).

\bibitem[{\citenamefont{Ghosh and Mandel}(1987)}]{GHOS87}
\bibinfo{author}{\bibfnamefont{R.}~\bibnamefont{Ghosh}} \bibnamefont{and}
  \bibinfo{author}{\bibfnamefont{L.}~\bibnamefont{Mandel}},
  \bibinfo{journal}{Phys. Rev. Lett.} \textbf{\bibinfo{volume}{59}},
  \bibinfo{pages}{1903} (\bibinfo{year}{1987}).

\bibitem[{\citenamefont{Yasuda and Shimizu}(1996)}]{YASU96}
\bibinfo{author}{\bibfnamefont{M.}~\bibnamefont{Yasuda}} \bibnamefont{and}
  \bibinfo{author}{\bibfnamefont{F.}~\bibnamefont{Shimizu}},
  \bibinfo{journal}{Phys. Rev. Lett.} \textbf{\bibinfo{volume}{77}},
  \bibinfo{pages}{3090 } (\bibinfo{year}{1996}).

\bibitem[{\citenamefont{Jeltes et~al.}(2007)\citenamefont{Jeltes, McNamara,
  Hogervorst, Vassen, Krachmalnicoff, Schellekens, Perrin, Chang, Boiron,
  Aspect et~al.}}]{JELT07}
\bibinfo{author}{\bibfnamefont{T.}~\bibnamefont{Jeltes}},
  \bibinfo{author}{\bibfnamefont{J.~M.} \bibnamefont{McNamara}},
  \bibinfo{author}{\bibfnamefont{W.}~\bibnamefont{Hogervorst}},
  \bibinfo{author}{\bibfnamefont{W.}~\bibnamefont{Vassen}},
  \bibinfo{author}{\bibfnamefont{V.}~\bibnamefont{Krachmalnicoff}},
  \bibinfo{author}{\bibfnamefont{M.}~\bibnamefont{Schellekens}},
  \bibinfo{author}{\bibfnamefont{A.}~\bibnamefont{Perrin}},
  \bibinfo{author}{\bibfnamefont{H.}~\bibnamefont{Chang}},
  \bibinfo{author}{\bibfnamefont{D.}~\bibnamefont{Boiron}},
  \bibinfo{author}{\bibfnamefont{A.}~\bibnamefont{Aspect}},
  \bibnamefont{et~al.}, \bibinfo{journal}{Nature}
  \textbf{\bibinfo{volume}{445}}, \bibinfo{pages}{402 } (\bibinfo{year}{2007}).

\bibitem[{\citenamefont{{\"{O}}ttl et~al.}(2005)\citenamefont{{\"{O}}ttl,
  Ritter, K{\"{o}}hl, and Esslinger}}]{OTTL05}
\bibinfo{author}{\bibfnamefont{A.}~\bibnamefont{{\"{O}}ttl}},
  \bibinfo{author}{\bibfnamefont{S.}~\bibnamefont{Ritter}},
  \bibinfo{author}{\bibfnamefont{M.}~\bibnamefont{K{\"{o}}hl}},
  \bibnamefont{and}
  \bibinfo{author}{\bibfnamefont{T.}~\bibnamefont{Esslinger}},
  \bibinfo{journal}{Phys. Rev. Lett.} \textbf{\bibinfo{volume}{95}},
  \bibinfo{pages}{090404} (\bibinfo{year}{2005}).

\bibitem[{\citenamefont{Schellekens et~al.}(2005)\citenamefont{Schellekens,
  Hoppeler, Perrin, J.V.ianaGomes, Boiron, Aspect, and Westbrook}}]{SCHE05}
\bibinfo{author}{\bibfnamefont{M.}~\bibnamefont{Schellekens}},
  \bibinfo{author}{\bibfnamefont{R.}~\bibnamefont{Hoppeler}},
  \bibinfo{author}{\bibfnamefont{A.}~\bibnamefont{Perrin}},
  \bibinfo{author}{\bibnamefont{J.V.ianaGomes}},
  \bibinfo{author}{\bibfnamefont{D.}~\bibnamefont{Boiron}},
  \bibinfo{author}{\bibfnamefont{A.}~\bibnamefont{Aspect}}, \bibnamefont{and}
  \bibinfo{author}{\bibfnamefont{C.}~\bibnamefont{Westbrook}},
  \bibinfo{journal}{Science} \textbf{\bibinfo{volume}{310}},
  \bibinfo{pages}{648 } (\bibinfo{year}{2005}).

\bibitem[{\citenamefont{Oliver et~al.}(1999)\citenamefont{Oliver, Kim, Liu, and
  Yamamoto}}]{OLIV99}
\bibinfo{author}{\bibfnamefont{W.}~\bibnamefont{Oliver}},
  \bibinfo{author}{\bibfnamefont{J.}~\bibnamefont{Kim}},
  \bibinfo{author}{\bibfnamefont{R.}~\bibnamefont{Liu}}, \bibnamefont{and}
  \bibinfo{author}{\bibfnamefont{Y.}~\bibnamefont{Yamamoto}},
  \bibinfo{journal}{Science} \textbf{\bibinfo{volume}{284}},
  \bibinfo{pages}{299 } (\bibinfo{year}{1999}).

\bibitem[{\citenamefont{Kiesel et~al.}(2002)\citenamefont{Kiesel, Renz, and
  Hasselbach}}]{KIES02}
\bibinfo{author}{\bibfnamefont{H.}~\bibnamefont{Kiesel}},
  \bibinfo{author}{\bibfnamefont{A.}~\bibnamefont{Renz}}, \bibnamefont{and}
  \bibinfo{author}{\bibfnamefont{F.}~\bibnamefont{Hasselbach}},
  \bibinfo{journal}{Nature} \textbf{\bibinfo{volume}{418}}, \bibinfo{pages}{392
  } (\bibinfo{year}{2002}).

\bibitem[{\citenamefont{{De Raedt} et~al.}(2005{\natexlab{a}})\citenamefont{{De
  Raedt}, {De Raedt}, and Michielsen}}]{RAED05d}
\bibinfo{author}{\bibfnamefont{H.}~\bibnamefont{{De Raedt}}},
  \bibinfo{author}{\bibfnamefont{K.}~\bibnamefont{{De Raedt}}},
  \bibnamefont{and}
  \bibinfo{author}{\bibfnamefont{K.}~\bibnamefont{Michielsen}},
  \bibinfo{journal}{Europhys. Lett.} \textbf{\bibinfo{volume}{69}},
  \bibinfo{pages}{861 } (\bibinfo{year}{2005}{\natexlab{a}}).

\bibitem[{\citenamefont{{De Raedt} et~al.}(2005{\natexlab{b}})\citenamefont{{De
  Raedt}, {De Raedt}, and Michielsen}}]{RAED05b}
\bibinfo{author}{\bibfnamefont{K.}~\bibnamefont{{De Raedt}}},
  \bibinfo{author}{\bibfnamefont{H.}~\bibnamefont{{De Raedt}}},
  \bibnamefont{and}
  \bibinfo{author}{\bibfnamefont{K.}~\bibnamefont{Michielsen}},
  \bibinfo{journal}{Comp. Phys. Comm.} \textbf{\bibinfo{volume}{171}},
  \bibinfo{pages}{19 } (\bibinfo{year}{2005}{\natexlab{b}}).

\bibitem[{\citenamefont{{Michielsen} et~al.}(2011)\citenamefont{{Michielsen},
  Jin, and {De Raedt}}}]{MICH11a}
\bibinfo{author}{\bibfnamefont{K.}~\bibnamefont{{Michielsen}}},
  \bibinfo{author}{\bibfnamefont{F.}~\bibnamefont{Jin}}, \bibnamefont{and}
  \bibinfo{author}{\bibfnamefont{H.}~\bibnamefont{{De Raedt}}},
  \bibinfo{journal}{J. Comp. Theor. Nanosci.} \textbf{\bibinfo{volume}{8}},
  \bibinfo{pages}{1052 } (\bibinfo{year}{2011}).

\bibitem[{\citenamefont{{Zhao} et~al.}(2008{\natexlab{a}})\citenamefont{{Zhao},
  {Yuan}, {De Raedt}, and Michielsen}}]{ZHAO08b}
\bibinfo{author}{\bibfnamefont{S.}~\bibnamefont{{Zhao}}},
  \bibinfo{author}{\bibfnamefont{S.}~\bibnamefont{{Yuan}}},
  \bibinfo{author}{\bibfnamefont{H.}~\bibnamefont{{De Raedt}}},
  \bibnamefont{and}
  \bibinfo{author}{\bibfnamefont{K.}~\bibnamefont{Michielsen}},
  \bibinfo{journal}{Europhys. Lett.} \textbf{\bibinfo{volume}{82}},
  \bibinfo{pages}{40004} (\bibinfo{year}{2008}{\natexlab{a}}).

\bibitem[{\citenamefont{Michielsen et~al.}(2010)\citenamefont{Michielsen, Yuan,
  Zhao, Jin, and {De Raedt}}}]{MICH10a}
\bibinfo{author}{\bibfnamefont{K.}~\bibnamefont{Michielsen}},
  \bibinfo{author}{\bibfnamefont{S.}~\bibnamefont{Yuan}},
  \bibinfo{author}{\bibfnamefont{S.}~\bibnamefont{Zhao}},
  \bibinfo{author}{\bibfnamefont{F.}~\bibnamefont{Jin}}, \bibnamefont{and}
  \bibinfo{author}{\bibfnamefont{H.}~\bibnamefont{{De Raedt}}},
  \bibinfo{journal}{Physica E} \textbf{\bibinfo{volume}{42}},
  \bibinfo{pages}{348 } (\bibinfo{year}{2010}).

\bibitem[{\citenamefont{{Jin} et~al.}(2010{\natexlab{a}})\citenamefont{{Jin},
  {Zhao}, {Yuan}, {De Raedt}, and {Michielsen}}}]{JIN10c}
\bibinfo{author}{\bibfnamefont{F.}~\bibnamefont{{Jin}}},
  \bibinfo{author}{\bibfnamefont{S.}~\bibnamefont{{Zhao}}},
  \bibinfo{author}{\bibfnamefont{S.}~\bibnamefont{{Yuan}}},
  \bibinfo{author}{\bibfnamefont{H.}~\bibnamefont{{De Raedt}}},
  \bibnamefont{and}
  \bibinfo{author}{\bibfnamefont{K.}~\bibnamefont{{Michielsen}}},
  \bibinfo{journal}{J. Comp. Theor. Nanosci.} \textbf{\bibinfo{volume}{7}},
  \bibinfo{pages}{1771} (\bibinfo{year}{2010}{\natexlab{a}}).

\bibitem[{\citenamefont{{Jin} et~al.}(2010{\natexlab{b}})\citenamefont{{Jin},
  {Yuan}, {De Raedt}, {Michielsen}, and Miyashita}}]{JIN10b}
\bibinfo{author}{\bibfnamefont{F.}~\bibnamefont{{Jin}}},
  \bibinfo{author}{\bibfnamefont{S.}~\bibnamefont{{Yuan}}},
  \bibinfo{author}{\bibfnamefont{H.}~\bibnamefont{{De Raedt}}},
  \bibinfo{author}{\bibfnamefont{K.}~\bibnamefont{{Michielsen}}},
  \bibnamefont{and}
  \bibinfo{author}{\bibfnamefont{S.}~\bibnamefont{Miyashita}},
  \bibinfo{journal}{J. Phys. Soc. Jpn.} \textbf{\bibinfo{volume}{79}},
  \bibinfo{pages}{074401} (\bibinfo{year}{2010}{\natexlab{b}}).

\bibitem[{\citenamefont{{Jin} et~al.}(2010{\natexlab{c}})\citenamefont{{Jin},
  {De Raedt}, and {Michielsen}}}]{JIN10a}
\bibinfo{author}{\bibfnamefont{F.}~\bibnamefont{{Jin}}},
  \bibinfo{author}{\bibfnamefont{H.}~\bibnamefont{{De Raedt}}},
  \bibnamefont{and}
  \bibinfo{author}{\bibfnamefont{K.}~\bibnamefont{{Michielsen}}},
  \bibinfo{journal}{Commun. Comput. Phys.} \textbf{\bibinfo{volume}{7}},
  \bibinfo{pages}{813 } (\bibinfo{year}{2010}{\natexlab{c}}).

\bibitem[{\citenamefont{{Michielsen} et~al.}(2012)\citenamefont{{Michielsen},
  {Lippert}, {Richter}, {Barbara}, {Miyashita}, and {De Raedt}}}]{MICH12a}
\bibinfo{author}{\bibfnamefont{K.}~\bibnamefont{{Michielsen}}},
  \bibinfo{author}{\bibfnamefont{T.}~\bibnamefont{{Lippert}}},
  \bibinfo{author}{\bibfnamefont{M.}~\bibnamefont{{Richter}}},
  \bibinfo{author}{\bibfnamefont{B.}~\bibnamefont{{Barbara}}},
  \bibinfo{author}{\bibfnamefont{S.}~\bibnamefont{{Miyashita}}},
  \bibnamefont{and} \bibinfo{author}{\bibfnamefont{H.}~\bibnamefont{{De
  Raedt}}}, \bibinfo{journal}{J. Phys. Soc. Jpn.}
  \textbf{\bibinfo{volume}{81}}, \bibinfo{pages}{034001}
  (\bibinfo{year}{2012}).

\bibitem[{\citenamefont{Grangier et~al.}(1986)\citenamefont{Grangier, Roger,
  and Aspect}}]{GRAN86}
\bibinfo{author}{\bibfnamefont{P.}~\bibnamefont{Grangier}},
  \bibinfo{author}{\bibfnamefont{G.}~\bibnamefont{Roger}}, \bibnamefont{and}
  \bibinfo{author}{\bibfnamefont{A.}~\bibnamefont{Aspect}},
  \bibinfo{journal}{Europhys. Lett.} \textbf{\bibinfo{volume}{1}},
  \bibinfo{pages}{173} (\bibinfo{year}{1986}).

\bibitem[{\citenamefont{Jacques et~al.}(2007)\citenamefont{Jacques, Wu,
  Grosshans, Treussart, Grangier, Aspect, and Roch}}]{JACQ07}
\bibinfo{author}{\bibfnamefont{V.}~\bibnamefont{Jacques}},
  \bibinfo{author}{\bibfnamefont{E.}~\bibnamefont{Wu}},
  \bibinfo{author}{\bibfnamefont{F.}~\bibnamefont{Grosshans}},
  \bibinfo{author}{\bibfnamefont{F.}~\bibnamefont{Treussart}},
  \bibinfo{author}{\bibfnamefont{P.}~\bibnamefont{Grangier}},
  \bibinfo{author}{\bibfnamefont{A.}~\bibnamefont{Aspect}}, \bibnamefont{and}
  \bibinfo{author}{\bibfnamefont{J.-F.} \bibnamefont{Roch}},
  \bibinfo{journal}{Science} \textbf{\bibinfo{volume}{315}},
  \bibinfo{pages}{966} (\bibinfo{year}{2007}).

\bibitem[{\citenamefont{Schwindt et~al.}(1999)\citenamefont{Schwindt, Kwiat,
  and Englert}}]{SCHW99}
\bibinfo{author}{\bibfnamefont{P.~D.~D.} \bibnamefont{Schwindt}},
  \bibinfo{author}{\bibfnamefont{P.~G.} \bibnamefont{Kwiat}}, \bibnamefont{and}
  \bibinfo{author}{\bibfnamefont{B.-G.} \bibnamefont{Englert}},
  \bibinfo{journal}{Phys. Rev. A} \textbf{\bibinfo{volume}{60}},
  \bibinfo{pages}{4285 } (\bibinfo{year}{1999}).

\bibitem[{\citenamefont{Jacques et~al.}(2005)\citenamefont{Jacques, Wu, Toury,
  Treussart, Aspect, Grangier, and Roch}}]{JACQ05}
\bibinfo{author}{\bibfnamefont{V.}~\bibnamefont{Jacques}},
  \bibinfo{author}{\bibfnamefont{E.}~\bibnamefont{Wu}},
  \bibinfo{author}{\bibfnamefont{T.}~\bibnamefont{Toury}},
  \bibinfo{author}{\bibfnamefont{F.}~\bibnamefont{Treussart}},
  \bibinfo{author}{\bibfnamefont{A.}~\bibnamefont{Aspect}},
  \bibinfo{author}{\bibfnamefont{P.}~\bibnamefont{Grangier}}, \bibnamefont{and}
  \bibinfo{author}{\bibfnamefont{J.-F.} \bibnamefont{Roch}},
  \bibinfo{journal}{Eur. Phys. J. D} \textbf{\bibinfo{volume}{35}},
  \bibinfo{pages}{561} (\bibinfo{year}{2005}).

\bibitem[{\citenamefont{{Zhao} and {De Raedt}}(2008)}]{ZHAO08a}
\bibinfo{author}{\bibfnamefont{S.}~\bibnamefont{{Zhao}}} \bibnamefont{and}
  \bibinfo{author}{\bibfnamefont{H.}~\bibnamefont{{De Raedt}}},
  \bibinfo{journal}{J. Comp. Theor. Nanosci.} \textbf{\bibinfo{volume}{5}},
  \bibinfo{pages}{490 } (\bibinfo{year}{2008}).

\bibitem[{\citenamefont{Agafonov et~al.}(2008)\citenamefont{Agafonov, Chekhova,
  Iskhakov, and Penin}}]{AGAF08}
\bibinfo{author}{\bibfnamefont{I.~N.} \bibnamefont{Agafonov}},
  \bibinfo{author}{\bibfnamefont{M.~V.} \bibnamefont{Chekhova}},
  \bibinfo{author}{\bibfnamefont{T.~S.} \bibnamefont{Iskhakov}},
  \bibnamefont{and} \bibinfo{author}{\bibfnamefont{A.~N.} \bibnamefont{Penin}},
  \bibinfo{journal}{Phys. Rev. A} \textbf{\bibinfo{volume}{77}},
  \bibinfo{pages}{053801} (\bibinfo{year}{2008}).

\bibitem[{\citenamefont{{De Raedt} et~al.}(2005{\natexlab{c}})\citenamefont{{De
  Raedt}, {De Raedt}, and Michielsen}}]{RAED05c}
\bibinfo{author}{\bibfnamefont{H.}~\bibnamefont{{De Raedt}}},
  \bibinfo{author}{\bibfnamefont{K.}~\bibnamefont{{De Raedt}}},
  \bibnamefont{and}
  \bibinfo{author}{\bibfnamefont{K.}~\bibnamefont{Michielsen}},
  \bibinfo{journal}{J. Phys. Soc. Jpn. Suppl.} \textbf{\bibinfo{volume}{76}},
  \bibinfo{pages}{16 } (\bibinfo{year}{2005}{\natexlab{c}}).

\bibitem[{\citenamefont{Michielsen et~al.}(2005)\citenamefont{Michielsen, {De
  Raedt}, and {De Raedt}}}]{MICH05}
\bibinfo{author}{\bibfnamefont{K.}~\bibnamefont{Michielsen}},
  \bibinfo{author}{\bibfnamefont{K.}~\bibnamefont{{De Raedt}}},
  \bibnamefont{and} \bibinfo{author}{\bibfnamefont{H.}~\bibnamefont{{De
  Raedt}}}, \bibinfo{journal}{J. Comput. Theor. Nanosci.}
  \textbf{\bibinfo{volume}{2}}, \bibinfo{pages}{227 } (\bibinfo{year}{2005}).

\bibitem[{\citenamefont{Aspect et~al.}(1982{\natexlab{a}})\citenamefont{Aspect,
  Grangier, and Roger}}]{ASPE82a}
\bibinfo{author}{\bibfnamefont{A.}~\bibnamefont{Aspect}},
  \bibinfo{author}{\bibfnamefont{P.}~\bibnamefont{Grangier}}, \bibnamefont{and}
  \bibinfo{author}{\bibfnamefont{G.}~\bibnamefont{Roger}},
  \bibinfo{journal}{Phys. Rev. Lett.} \textbf{\bibinfo{volume}{49}},
  \bibinfo{pages}{91 } (\bibinfo{year}{1982}{\natexlab{a}}).

\bibitem[{\citenamefont{Aspect et~al.}(1982{\natexlab{b}})\citenamefont{Aspect,
  Dalibard, and Roger}}]{ASPE82b}
\bibinfo{author}{\bibfnamefont{A.}~\bibnamefont{Aspect}},
  \bibinfo{author}{\bibfnamefont{J.}~\bibnamefont{Dalibard}}, \bibnamefont{and}
  \bibinfo{author}{\bibfnamefont{G.}~\bibnamefont{Roger}},
  \bibinfo{journal}{Phys. Rev. Lett.} \textbf{\bibinfo{volume}{49}},
  \bibinfo{pages}{1804 } (\bibinfo{year}{1982}{\natexlab{b}}).

\bibitem[{\citenamefont{Weihs et~al.}(1998)\citenamefont{Weihs, Jennewein,
  Simon, Weinfurther, and Zeilinger}}]{WEIH98}
\bibinfo{author}{\bibfnamefont{G.}~\bibnamefont{Weihs}},
  \bibinfo{author}{\bibfnamefont{T.}~\bibnamefont{Jennewein}},
  \bibinfo{author}{\bibfnamefont{C.}~\bibnamefont{Simon}},
  \bibinfo{author}{\bibfnamefont{H.}~\bibnamefont{Weinfurther}},
  \bibnamefont{and}
  \bibinfo{author}{\bibfnamefont{A.}~\bibnamefont{Zeilinger}},
  \bibinfo{journal}{Phys. Rev. Lett.} \textbf{\bibinfo{volume}{81}},
  \bibinfo{pages}{5039 } (\bibinfo{year}{1998}).

\bibitem[{\citenamefont{{De Raedt} et~al.}(2006)\citenamefont{{De Raedt},
  Keimpema, {De Raedt}, Michielsen, and Miyashita}}]{RAED06c}
\bibinfo{author}{\bibfnamefont{K.}~\bibnamefont{{De Raedt}}},
  \bibinfo{author}{\bibfnamefont{K.}~\bibnamefont{Keimpema}},
  \bibinfo{author}{\bibfnamefont{H.}~\bibnamefont{{De Raedt}}},
  \bibinfo{author}{\bibfnamefont{K.}~\bibnamefont{Michielsen}},
  \bibnamefont{and}
  \bibinfo{author}{\bibfnamefont{S.}~\bibnamefont{Miyashita}},
  \bibinfo{journal}{Euro. Phys. J. B} \textbf{\bibinfo{volume}{53}},
  \bibinfo{pages}{139 } (\bibinfo{year}{2006}).

\bibitem[{\citenamefont{{De Raedt} et~al.}(2007{\natexlab{a}})\citenamefont{{De
  Raedt}, {De Raedt}, Michielsen, Keimpema, and Miyashita}}]{RAED07a}
\bibinfo{author}{\bibfnamefont{H.}~\bibnamefont{{De Raedt}}},
  \bibinfo{author}{\bibfnamefont{K.}~\bibnamefont{{De Raedt}}},
  \bibinfo{author}{\bibfnamefont{K.}~\bibnamefont{Michielsen}},
  \bibinfo{author}{\bibfnamefont{K.}~\bibnamefont{Keimpema}}, \bibnamefont{and}
  \bibinfo{author}{\bibfnamefont{S.}~\bibnamefont{Miyashita}},
  \bibinfo{journal}{J. Phys. Soc. Jpn.} \textbf{\bibinfo{volume}{76}},
  \bibinfo{pages}{104005} (\bibinfo{year}{2007}{\natexlab{a}}).

\bibitem[{\citenamefont{{De Raedt} et~al.}(2007{\natexlab{b}})\citenamefont{{De
  Raedt}, {De Raedt}, and Michielsen}}]{RAED07b}
\bibinfo{author}{\bibfnamefont{K.}~\bibnamefont{{De Raedt}}},
  \bibinfo{author}{\bibfnamefont{H.}~\bibnamefont{{De Raedt}}},
  \bibnamefont{and}
  \bibinfo{author}{\bibfnamefont{K.}~\bibnamefont{Michielsen}},
  \bibinfo{journal}{Comp. Phys. Comm.} \textbf{\bibinfo{volume}{176}},
  \bibinfo{pages}{642 } (\bibinfo{year}{2007}{\natexlab{b}}).

\bibitem[{\citenamefont{{De Raedt} et~al.}(2007{\natexlab{c}})\citenamefont{{De
  Raedt}, {De Raedt}, Michielsen, Keimpema, and Miyashita}}]{RAED07c}
\bibinfo{author}{\bibfnamefont{H.}~\bibnamefont{{De Raedt}}},
  \bibinfo{author}{\bibfnamefont{K.}~\bibnamefont{{De Raedt}}},
  \bibinfo{author}{\bibfnamefont{K.}~\bibnamefont{Michielsen}},
  \bibinfo{author}{\bibfnamefont{K.}~\bibnamefont{Keimpema}}, \bibnamefont{and}
  \bibinfo{author}{\bibfnamefont{S.}~\bibnamefont{Miyashita}},
  \bibinfo{journal}{J. Comp. Theor. Nanosci.} \textbf{\bibinfo{volume}{4}},
  \bibinfo{pages}{957 } (\bibinfo{year}{2007}{\natexlab{c}}).

\bibitem[{\citenamefont{{De Raedt} et~al.}(2007{\natexlab{d}})\citenamefont{{De
  Raedt}, Michielsen, Miyashita, and Keimpema}}]{RAED07d}
\bibinfo{author}{\bibfnamefont{H.}~\bibnamefont{{De Raedt}}},
  \bibinfo{author}{\bibfnamefont{K.}~\bibnamefont{Michielsen}},
  \bibinfo{author}{\bibfnamefont{S.}~\bibnamefont{Miyashita}},
  \bibnamefont{and} \bibinfo{author}{\bibfnamefont{K.}~\bibnamefont{Keimpema}},
  \bibinfo{journal}{Euro. Phys. J. B} \textbf{\bibinfo{volume}{58}},
  \bibinfo{pages}{55 } (\bibinfo{year}{2007}{\natexlab{d}}).

\bibitem[{\citenamefont{{Zhao} et~al.}(2008{\natexlab{b}})\citenamefont{{Zhao},
  {De Raedt}, and Michielsen}}]{ZHAO08}
\bibinfo{author}{\bibfnamefont{S.}~\bibnamefont{{Zhao}}},
  \bibinfo{author}{\bibfnamefont{H.}~\bibnamefont{{De Raedt}}},
  \bibnamefont{and}
  \bibinfo{author}{\bibfnamefont{K.}~\bibnamefont{Michielsen}},
  \bibinfo{journal}{Found. of Phys.} \textbf{\bibinfo{volume}{38}},
  \bibinfo{pages}{322 } (\bibinfo{year}{2008}{\natexlab{b}}).

\bibitem[{\citenamefont{{Trieu} et~al.}(2011)\citenamefont{{Trieu},
  {Michielsen}, and {De Raedt}}}]{TRIE11}
\bibinfo{author}{\bibfnamefont{B.}~\bibnamefont{{Trieu}}},
  \bibinfo{author}{\bibfnamefont{K.}~\bibnamefont{{Michielsen}}},
  \bibnamefont{and} \bibinfo{author}{\bibfnamefont{H.}~\bibnamefont{{De
  Raedt}}}, \bibinfo{journal}{Comp. Phys. Comm.}
  \textbf{\bibinfo{volume}{182}}, \bibinfo{pages}{726} (\bibinfo{year}{2011}).

\bibitem[{\citenamefont{Pfleegor and Mandel}(1967)}]{PFlE67}
\bibinfo{author}{\bibfnamefont{R.}~\bibnamefont{Pfleegor}} \bibnamefont{and}
  \bibinfo{author}{\bibfnamefont{L.}~\bibnamefont{Mandel}},
  \bibinfo{journal}{Phys. Rev.} \textbf{\bibinfo{volume}{159}},
  \bibinfo{pages}{1084 } (\bibinfo{year}{1967}).

\bibitem[{\citenamefont{Hadfield}(2009)}]{HADF09}
\bibinfo{author}{\bibfnamefont{R.~H.} \bibnamefont{Hadfield}},
  \bibinfo{journal}{Nature Photonics} \textbf{\bibinfo{volume}{3}},
  \bibinfo{pages}{696 } (\bibinfo{year}{2009}).

\bibitem[{\citenamefont{Garrison and Chiao}(2009)}]{GARR09}
\bibinfo{author}{\bibfnamefont{J.~C.} \bibnamefont{Garrison}} \bibnamefont{and}
  \bibinfo{author}{\bibfnamefont{R.~Y.} \bibnamefont{Chiao}},
  \emph{\bibinfo{title}{{Quantum Optics}}} (\bibinfo{publisher}{Oxford
  University Press}, \bibinfo{address}{Oxford}, \bibinfo{year}{2009}).

\bibitem[{\citenamefont{Home}(1997)}]{HOME97}
\bibinfo{author}{\bibfnamefont{D.}~\bibnamefont{Home}},
  \emph{\bibinfo{title}{{Conceptual Foundations of Quantum Physics}}}
  (\bibinfo{publisher}{Plenum Press}, \bibinfo{address}{New York},
  \bibinfo{year}{1997}).

\bibitem[{\citenamefont{Rubin and Shih}(1992)}]{RUBI92}
\bibinfo{author}{\bibfnamefont{M.}~\bibnamefont{Rubin}} \bibnamefont{and}
  \bibinfo{author}{\bibfnamefont{Y.~H.} \bibnamefont{Shih}},
  \bibinfo{journal}{Phys. Rev. A} \textbf{\bibinfo{volume}{45}},
  \bibinfo{pages}{8138 } (\bibinfo{year}{1992}).

\bibitem[{\citenamefont{Shih et~al.}(1993)\citenamefont{Shih, Sergienko, and
  Rubin}}]{SHIH93}
\bibinfo{author}{\bibfnamefont{Y.~H.} \bibnamefont{Shih}},
  \bibinfo{author}{\bibfnamefont{A.~V.} \bibnamefont{Sergienko}},
  \bibnamefont{and} \bibinfo{author}{\bibfnamefont{M.~H.} \bibnamefont{Rubin}},
  \bibinfo{journal}{Phys. Rev. A} \textbf{\bibinfo{volume}{47}},
  \bibinfo{pages}{1288 } (\bibinfo{year}{1993}).

\bibitem[{\citenamefont{Shih et~al.}(1994)\citenamefont{Shih, Sergienko, Rubin,
  Kiess, and Alley}}]{SHIH94}
\bibinfo{author}{\bibfnamefont{Y.~H.} \bibnamefont{Shih}},
  \bibinfo{author}{\bibfnamefont{A.~V.} \bibnamefont{Sergienko}},
  \bibinfo{author}{\bibfnamefont{M.~H.} \bibnamefont{Rubin}},
  \bibinfo{author}{\bibfnamefont{T.~E.} \bibnamefont{Kiess}}, \bibnamefont{and}
  \bibinfo{author}{\bibfnamefont{C.~O.} \bibnamefont{Alley}},
  \bibinfo{journal}{Phys. Re} \textbf{\bibinfo{volume}{49}},
  \bibinfo{pages}{4243 } (\bibinfo{year}{1994}).

\end{thebibliography}

\end{document}